\documentclass[twocolumn,floatfix,superscriptaddress,citeautoscript,10pt]{revtex4-1}

\usepackage{graphicx}
\usepackage{color}
\usepackage{siunitx}
\usepackage{float}
\usepackage{xcolor}
\newcommand{\siv}{SiV$^-$}
\newcommand{\sn}{Si$_3$N$_4$}
\newcommand{\tief}[1]{_\mathrm{#1}}
\newcommand{\gammac}{\Gamma_\mathrm{cav}}
\newcommand{\gammaf}{\Gamma_\mathrm{free}}

\DeclareSIUnit{\counts}{c}

\begin{document}

\title{Purcell-Enhanced Emission from Individual \siv{} Center in Nanodiamonds Coupled to a \sn{}-Based, Photonic Crystal Cavity}%
\author{Konstantin G. Fehler}
%\altaffiliation{Both authors contributed equally to this work.}
\affiliation{Institut f\"ur Quantenoptik, Universit\"at Ulm, D-89081 Ulm, Germany}
\affiliation{Center for Integrated Quantum Science and Technology (IQst), Ulm University, Albert-Einstein-Allee 11, D-89081 Ulm, Germany}
\author{Anna P. Ovvyan}
%\altaffiliation{Both authors contributed equally to this work.}
\affiliation{Institute of Physics and Center for Nanotechnology, University of Münster, D-48149 Münster, Germany}
\author{Lukas Antoniuk}
\affiliation{Institut f\"ur Quantenoptik, Universit\"at Ulm, D-89081 Ulm, Germany}
\author{Niklas Lettner}
\affiliation{Institut f\"ur Quantenoptik, Universit\"at Ulm, D-89081 Ulm, Germany}
\author{Nico Gruhler}
\affiliation{Institute of Physics and Center for Nanotechnology, University of Münster, D-48149 Münster, Germany}
\author{Valery A. Davydov}
\affiliation{L.F. Vereshchagin Institute for High Pressure Physics, Russian Academy of Sciences, Troitsk, Moscow 142190, Russia}
\author{Viatcheslav N. Agafonov}
\affiliation{GREMAN, UMR CNRS CEA 6157, Université F. Rabelais, 37200 Tours, France}
\author{Wolfram H. P. Pernice}
\affiliation{Institute of Physics and Center for Nanotechnology, University of Münster, D-48149 Münster, Germany}
\author{Alexander Kubanek}
\email{alexander.kubanek@uni-ulm.de}
\affiliation{Institut f\"ur Quantenoptik, Universit\"at Ulm, D-89081 Ulm, Germany}
\affiliation{Center for Integrated Quantum Science and Technology (IQst), Ulm University, Albert-Einstein-Allee 11, D-89081 Ulm, Germany}

\begin{abstract}
Hybrid quantum photonics combines classical photonics with quantum emitters in a postprocessing step. It facilitates to link ideal quantum light sources to optimized photonic platforms. Optical cavities enable to harness the Purcell-effect boosting the device efficiency. Here, we postprocess a free-standing, crossed-waveguide photonic crystal cavity based on \sn{} with \siv{} center in nanodiamonds. We develop a routine that holds the capability to optimize all degrees of freedom of the evanescent coupling term utilizing AFM nanomanipulation. After a few optimization cycles we resolve the fine-structure of individual \siv{} centers and achieve a Purcell enhancement of more than 4 on individual optical transitions, meaning that four out of five spontaneously emitted photons are channeled into the photonic device. Our work opens up new avenues to construct efficient quantum photonic devices. 
\end{abstract}

\maketitle
\section{Introduction}
Diamond is among the leading material platforms for spin-based photonic quantum technologies \cite{atature_material_2018, awschalom_quantum_2018}.
The negatively-charged silicon-vacancy center (\siv{} center) became one of the most promising color center in diamond due to strong zero phonon line (ZPL) emission, narrow inhomogeneous distribution and negligible spectral diffusion 
\cite{becker_coherence_2017, rogers_single_2019}
enabling two-photon interference from distinct \siv{} centers without the need of frequency tuning \cite{sipahigil_indistinguishable_2014}. The electronic spin coherence is limited by rapid phonon-mediated relaxation but can be improved by applying high strain \cite{sohn_controlling_2018} or by suppressing phonon-mediated relaxation at milli Kelvin temperatures \cite{sukachev_silicon_2017}. Recently, the deterministic polarisation of a small nuclear spin ensemble via dynamic nuclear polarization was demonstrated \cite{metsch_initialization_2019}. \\
Integrating the \siv{} center into on-chip photonics enables efficient spin-photon interface by Purcell-enhancement and scalable photonic networks. Classical fabrication methods for photonics platforms based on materials such as GaP, Si or \sn{} show low photon loss, design fexibility, standardization of the fabrication process, high throughput production or scalability to large-scale designs. It is therefore desirable to functionalize classical photonics for quantum applications.

Hybrid approaches pick up that challenge by combining quantum emitters with the most suitable photonics platforms. The post-processing step is an extraordinary challenge and can, for example, be realized by evanescent coupling. An idealized procedure utilizes preselected quantum emitters in a nanometersized host matrix to position the quantum light source with high accuracy in the interaction zone of the photonic device. Hybrid attempts based on color centers in diamond and high-refractive index photonics devices have been demonstrated in the past years 
\cite{fu_coupling_2008, englund_deterministic_2010, barclay_hybrid_2011, wolters_enhancement_2010}
 with challenges arising from weak evanescent coupling, high-background fluorescence or $Q$-factor degradation \cite{radulaski_nanodiamond_2019}. Reasonably large coupling was achieved between an ensemble of $\mathrm{NV}^{-}$ center in nanodiamonds (NDs) and the mode of a high-$Q$, free-standing photonic crystal cavity (PCC) in \sn{} where, at the same time, the background fluorescence was suppressed by $\sim$ \SI{-20}{\decibel}
in a crossed-waveguide pump-probe design \cite{fehler_efficient_2019}.\\ 
In this work we post-process a high-$Q$ photonic crystal cavity (PCC) based on \sn{} which was optimized for quantum photonics applications with \siv{} centers in NDs. We take advantage of bulk-like optical and coherence properties of \siv center in nanodiamonds \cite{jantzen_nanodiamonds_2016, hausler_preparing_2019}, and  high-precision nanomanipulation
tools \cite{schell_scanning_2011, rogers_single_2019} in order to access all degrees of freedom of the coupling term to the \siv center \cite{hausler_preparing_2019}. After a few optimization cycles on the evanescent coupling term we achieve a coherent coupling of the zero-phonon line (ZPL) to the mode of the PCC with a $\beta$ -factor of 0.44 and a Purcell factor of 0.79 averaged over an ensemble of \siv{} centers. After cooling the sample to liquid Helium temperatures we resolve the fine-structure of individual \siv{} centers and achieve a Purcell enhancement of more than 4 for individual optical transitions. The highly efficient coupling of individual atomic transitions to photonic circuits lays the foundation for quantum applications such as quantum networks \cite{wehner_quantum_2018} or on-chip Boson sampling \cite{spring_quantum_2013, li_nonblinking_2016, zhou_coherent_2017} based on hybrid quantum photonics.
\begin{figure*}[htp]
\includegraphics[width=2\columnwidth]{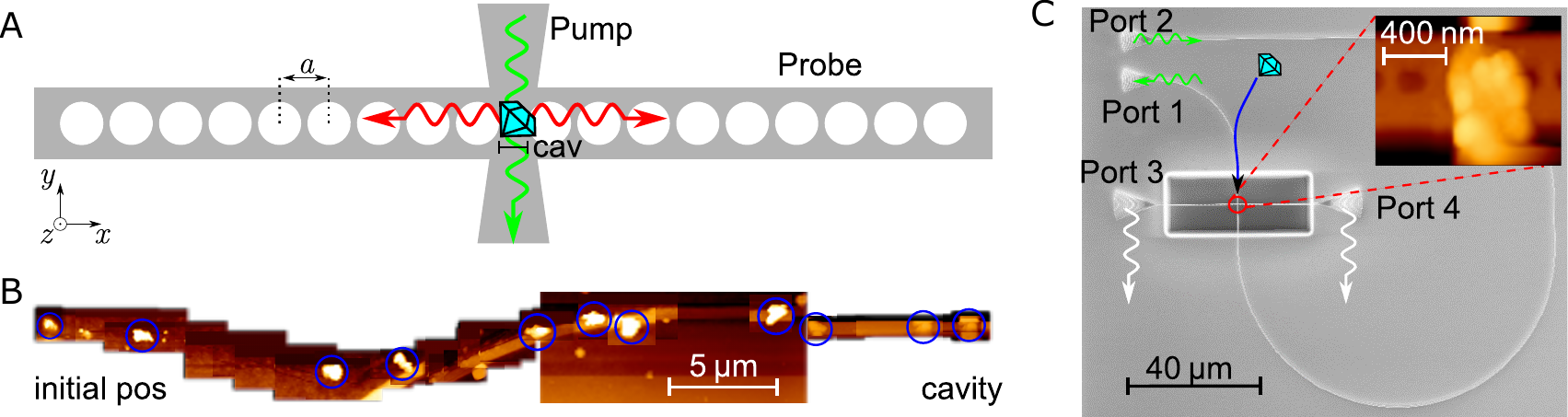}
\caption{Nanopositioning of ND within cavity mode. \textbf{A} The periodic change $a$ of the material on both sides of the probe waveguide lead to a photonic band gap, while the cavity is formed by the distance cav. The positioned ND can be evanescently excited through the pump waveguide. \textbf{B} AFM images during nanopositioning of \siv{} ensemble into PCC. The distance of the initial position was approximately $\SI{40}{\micro\meter}$ to the final position on top of the cavity. \textbf{C} SEM image of the PCC. The device has four grating couplers - two for coupling of green light (port 1 and 2) to off-resonantly excite the emitter in the interaction zone - and two grating couplers (port 3 and 4) for collecting the emission of the cavity system. The inset shows an AFM image of the ND ensemble placed in the interaction zone.}
\label{fig:afmpush_and_device}
\end{figure*}
\section{Results}
\subsection*{Photonic Crystal Cavity}

Our platform consists of a free-standing, crossed-waveguide PCC in \sn{} featuring low-loss transmission and high-mode confinement together with a minimized cross-talk between pump and probe waveguide ($\sim$ \SI{-20}{\decibel}). On-chip, off-resonant excitation and emission is spatially separated. While the pump waveguide is optimized for \SI{532}{\nano\meter}, the probe waveguide is optimized for \SI{740}{\nano\meter}, which matches the ZPL of the \siv{} center. Both waveguides are connected to grating couplers, which allow for out-of-plane excitation and readout. The on-chip excitation of the \siv{} centers is achieved by evanescent coupling to the pump waveguide. The probe waveguide hosts a 1D photonic crystal cavity with its modes superimposed to the pump volume. Each cavity mirror consist of $N=53$ holes with a period of $a=\SI{265}{\nano\meter}$. This periodic variation of permittivity forms a band gap in the visible range. A distance $\mathrm{cav}=\SI{232}{\nano\meter}$ between these two mirrors is inserted, which results in spectrally separated states inside the photonic band gap optimized by FDTD simulations. The nanophotonic circuit is post-processed with an \siv{} ensemble inside the crossing area of pump and probe beam (interaction zone), which is sketched in Fig. \ref{fig:afmpush_and_device} A.\\

\subsection*{Post-Processing}
The post-processing step for placing the ND within the interaction zone of the PCC follows the procedure described in Reference \cite{schell_scanning_2011,fehler_efficient_2019}.
A dispersion of NDs in water with incorporated \siv{} color centers is coated on the surface of the PCC chip. Suitable \siv{} center are preselected using a custom-build confocal microscope. Superimposing the confocal scan with an AFM image enables the precise localization of the \siv{} center host crystal. After the preselected ND is found, the AFM cantilever tip is used to push the ND in the interaction zone of the PCC.
Several steps of the positioning procedure are shown in Fig. \ref{fig:afmpush_and_device} B. The ND is pushed along a total distance of over $\SI{40}{\micro\meter}$.\\
The manipulation path is schematically drawn in the SEM image of the photonic device in Fig. \ref{fig:afmpush_and_device} C. The positioned ND is off-resonantly excited through port 1 and its emission into the cavity-waveguide can be read out at port 3 and 4. The small inset shows an AFM image of the positioned ND in the interaction zone.
A cross-talk measurement of the empty PCC probes the resonance modes. Therefore, a $\SI{532}{\nano\meter}$ laser with $\SI{0.6}{\milli\watt}$ is coupled into port 1 and the cavity signal is collected at port 3. The signal, arising from \sn{} background fluorescence, is shown in Fig \ref{fig:siv_warm} A (green). The occurring peaks correspond to the respective cavity modes, where the highest resonance at \SI{721}{\nano\meter} shows a quality factor of $Q=2260$ and the resonance near \SI{737.4}{\nano\meter} has a $Q$-factor of $Q=1000$.\\

\subsection*{Simulation of LDOS}
The design of the freestanding PCC is numerically optimized \textit{via} 3D FDTD simulations \cite{oskooi_meep_2010}. The PCC consists of two modulated Bragg mirrors with inserted cavity region in between. The periodicity of the holes ($a = \SI{265}{\nano\meter}$) is determined to match the bandgap region of the PCC to the investigated wavelength of the \siv{} ZPL. To achieve high $Q$-factors optimization on the cavity length and hole diameter was performed \cite{fehler_efficient_2019, akahane_high_2003}. 

For a maximum enhancement of the emitted light, the source need to be placed in the antinode of the electric field distribution of the resonance mode \cite{purcell_resonance_1946}. Thus, the position of the \siv{} embedded in a ND is carried out \textit{via} 3D FDTD simulations \cite{oskooi_meep_2010} in two steps. For the simulation a cube-shaped (\SI{200}{\nano\meter}) ND is placed on the cavity region.

\begin{figure*}[htp]
\includegraphics[width=2\columnwidth]{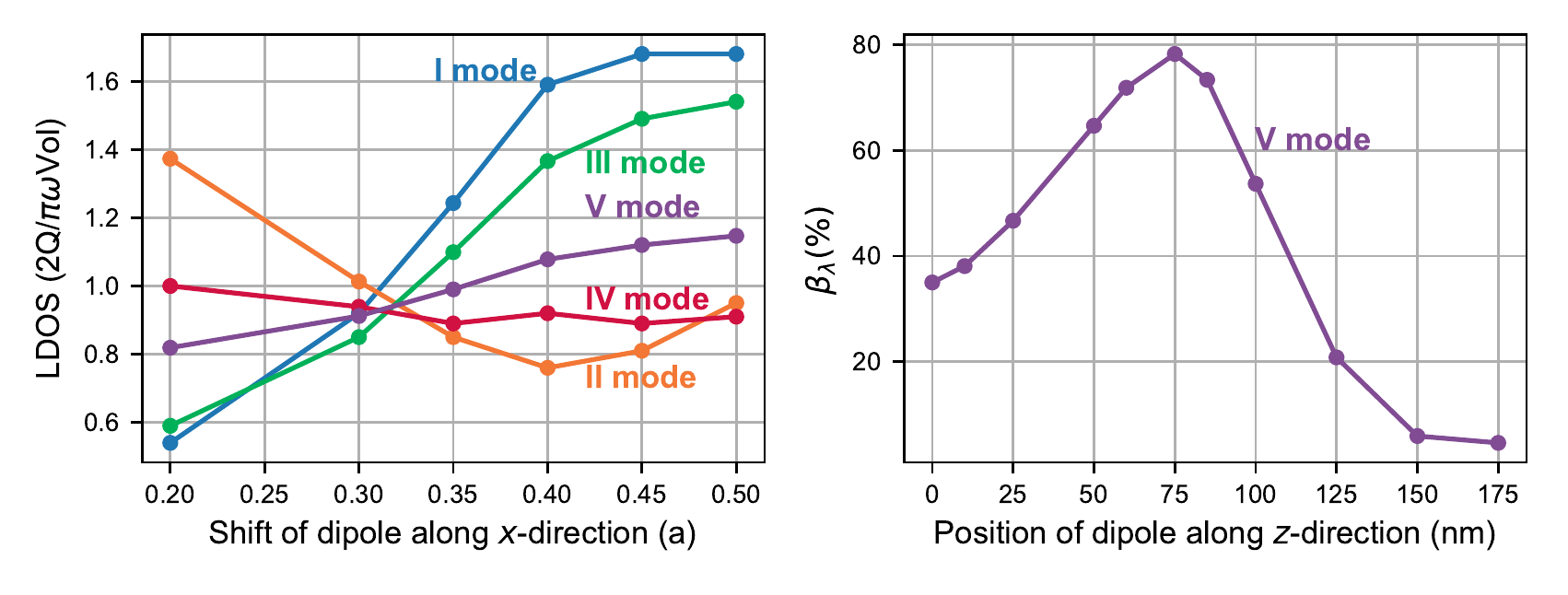}
\caption{Simulation of ND position on cavity interaction zone. \textbf{A} Simulated LDOS enhancement for different resonance modes of the emitted light from the dipole embedded in a \SI{200}{\nano\meter} cube-shaped ND, dependent on its longitudinal position weighted by the cavity periodicity $a$. The position of the ND is chosen according to its experimental position in Fig. \ref{fig:afmpush_and_device} C, while the dipole position is varied along the $x$-direction. \textbf{B} Simulated on-resonance coupling efficiency $\beta_\lambda$ into researched V-order resonance mode, dependent on the position of the dipole along the $z$-axis (keeping $y$ and $x$ position fixed at its optimum). The highest coupling efficiency is achieved for a displacement of \SI{60}{\nano\meter} above the cavity surface.}
\label{fig:simulations}
\end{figure*}

The coupling strength between the optical dipole transition and the cavity field mode depends, in particular, on the dipole orientation and the position with respect to the cavity field. Both parameters can be accessed by nanomanipulation as demonstrated in reference \cite{hausler_preparing_2019} without altering the optical properties of the \siv{} center. While the dipole orientation shows a cosine dependence the dipole position requires a more sophisticated study. The first optimization step is the enhancement of the Local Density of States (LDOS) dependent on the emitter position (see Fig. \ref{fig:simulations} A). The LDOS is proportional to the overlap integral between the electric field distribution of the resonance mode and the emitter \cite{taflove_photonics_2013}. According to the axes in Fig. \ref{fig:afmpush_and_device} A the dipole position, embedded in a ND placed on the cavity, is centered along $y$- and $z$-direction and varied along the longitudinal $x$-direction, since this component shows the strongest contribution to the convolution. The rotational orientation of the dipole is aligned to the PCC axis for highest coupling. The overlap integral of the dipole and the electric field follows a cosine dependence and is therefore one for the aligned case and zero for the orthogonal orientation. The normalized LDOS enhancement is calculated for each position of the dipole with respect to the field distribution of the mode. Therefore, the enhancement of the LDOS through the cavity is divided by the LDOS enhancement for a waveguide without cavity. The highest enhancement (see Fig. \ref{fig:simulations}A) for odd modes is achieved for a dipole shift of $\sim$ (0.4-0.45)$\,a$ from the symmetry plane of the center of the cavity, matching an antinode of the electric field distribution. For a reduction in simulation time, the number of segments in each Bragg-mirror was reduced to $N=18$, since the electric field distribution (position of minimum and maximum) of resonance modes changes in a minor way with varying number of segments.

The second part of the 3D FDTD simulations targeted the position optimization of the ND embedded emitter along the $z$-axis. Therefore, we quantify the coupling of the emitter to the cavity with the $\beta$-factor \cite{fehler_efficient_2019}, which gives the ratio of coupled spontaneous emission $\gammac$ and the total amount of spontaneously emitted photons to free space and in the cavity $\gammaf+\gammac$:
\begin{equation}
\beta = \frac{\gammac}{\gammac+\gammaf}
\label{eq:beta}
\end{equation}
The $\beta_\lambda$-factor (spectrally resolved $\beta$-factor \cite{fehler_efficient_2019}), correlating with LDOS enhancement, of the emission into the researched V-order resonance mode was examined. To consider the experimental degradation of the PCC due to the presence of the ND in the simulation, the number of segments in each Bragg-mirror was reduced. For this purpose, the number of holes is gradually adjusted from $N=53$ to 27 until the simulated $Q$-factor matches the experimentally examined $Q$-factor. The center of the ND was shifted according the to the position in the AFM image in Fig \ref{fig:afmpush_and_device} C. The $x$-position of the dipole inside the ND is set to its optimal according to Fig. \ref{fig:simulations} A, while the $z$-distance from the surface of the cavity is altered. The resulting coupling efficiencies for different $z$-positions are shown in Fig. \ref{fig:simulations} B. The maximum $\beta_\lambda$ value of \SI{78}{\percent} is reached for the source being located \SI{75}{\nano\meter} above the surface of the cavity. 

FDTD simulations are carried out to compute the Purcell enhancement \cite{purcell_resonance_1946} for the Vth-order resonance mode of the cross-bar PCCs with $N=27$ and $N=53$ holes, each with a \SI{200}{\nano\meter} cube-shaped ND crystal on the cavity region as described above. The resulting Purcell-factors read $F\tief{P}=20$ and $F\tief{P}=29$, respectively for a spectrally and polarization matched resonance mode and an emitter located at the antinode of the electric field.

\subsection*{Purcell-Enhanced Photon Emission}
\begin{figure}[htp]
\includegraphics[width=\columnwidth]{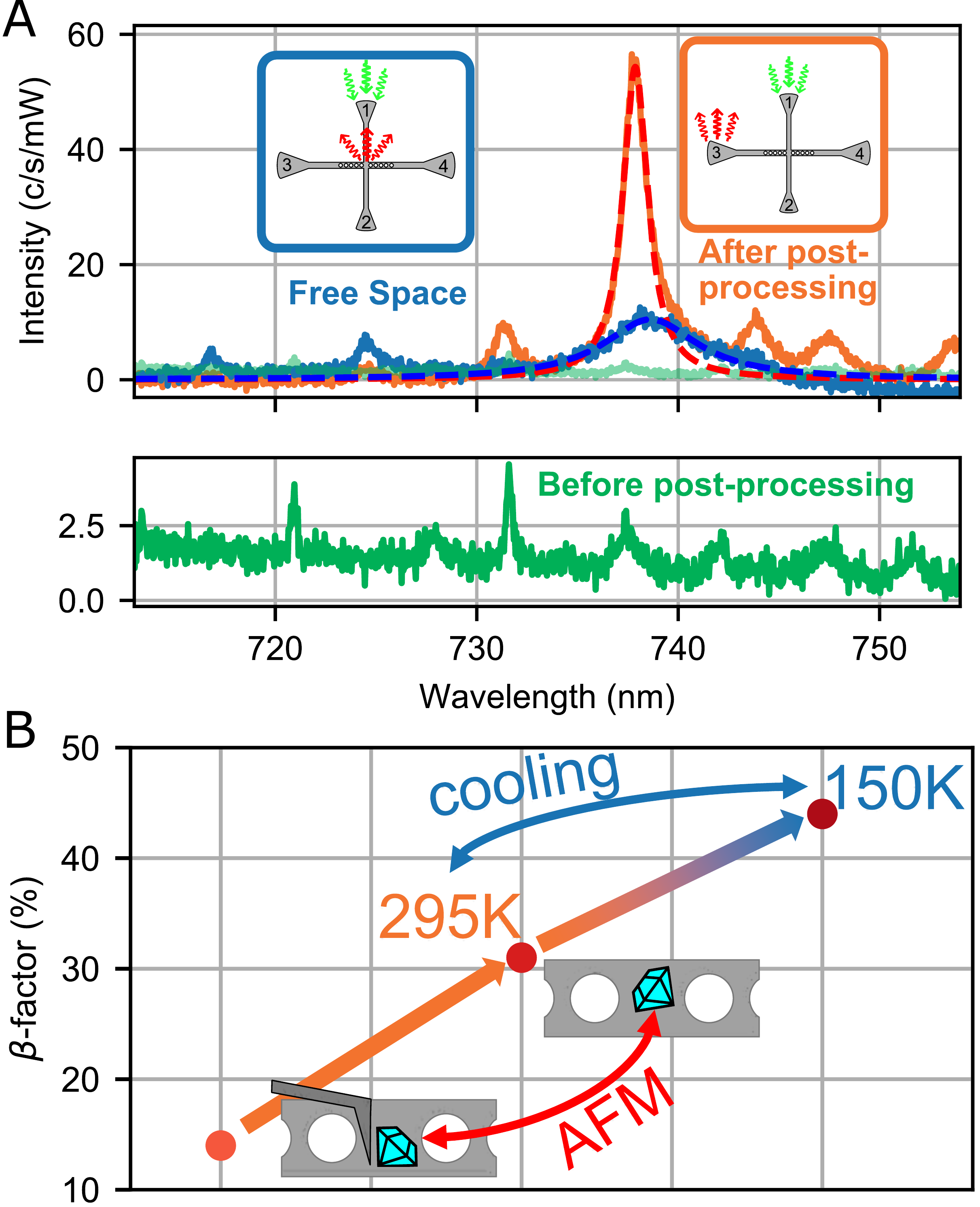}
\caption{Optimized emitter-cavity coupling. \textbf{A)} The cross-talk spectrum of the cavity (green) before the ND was placed, shows a resonance at \SI{737.4}{\nano\meter}. In blue the free space emission of the \siv{} ensemble for $\approx\SI{150}{\kelvin}$. The free space emission was collected in the center, while excited over port 1, as it can be seen in the small inset. The spectrum of the coupled cavity emission in orange shows a clear magnification of the emission rate for the resonance at \SI{737.9}{\nano\meter} (excited over port 1 and collected from port 3). A $\beta$-factor of 0.44 corresponding to a Purcell of 0.79 was calculated from the Lorentzian fits on the free space and the cavity-emitter emission spectra. \textbf{B} Coupling evolution for different tuning mechanism like positioning of the ND inside the interaction zone and temperature tuning. The highest ensemble coupling was achieved for Position 2 and \SI{150}{\kelvin}.}
\label{fig:siv_warm}
\end{figure}

\begin{figure*}
\includegraphics[width=2\columnwidth]{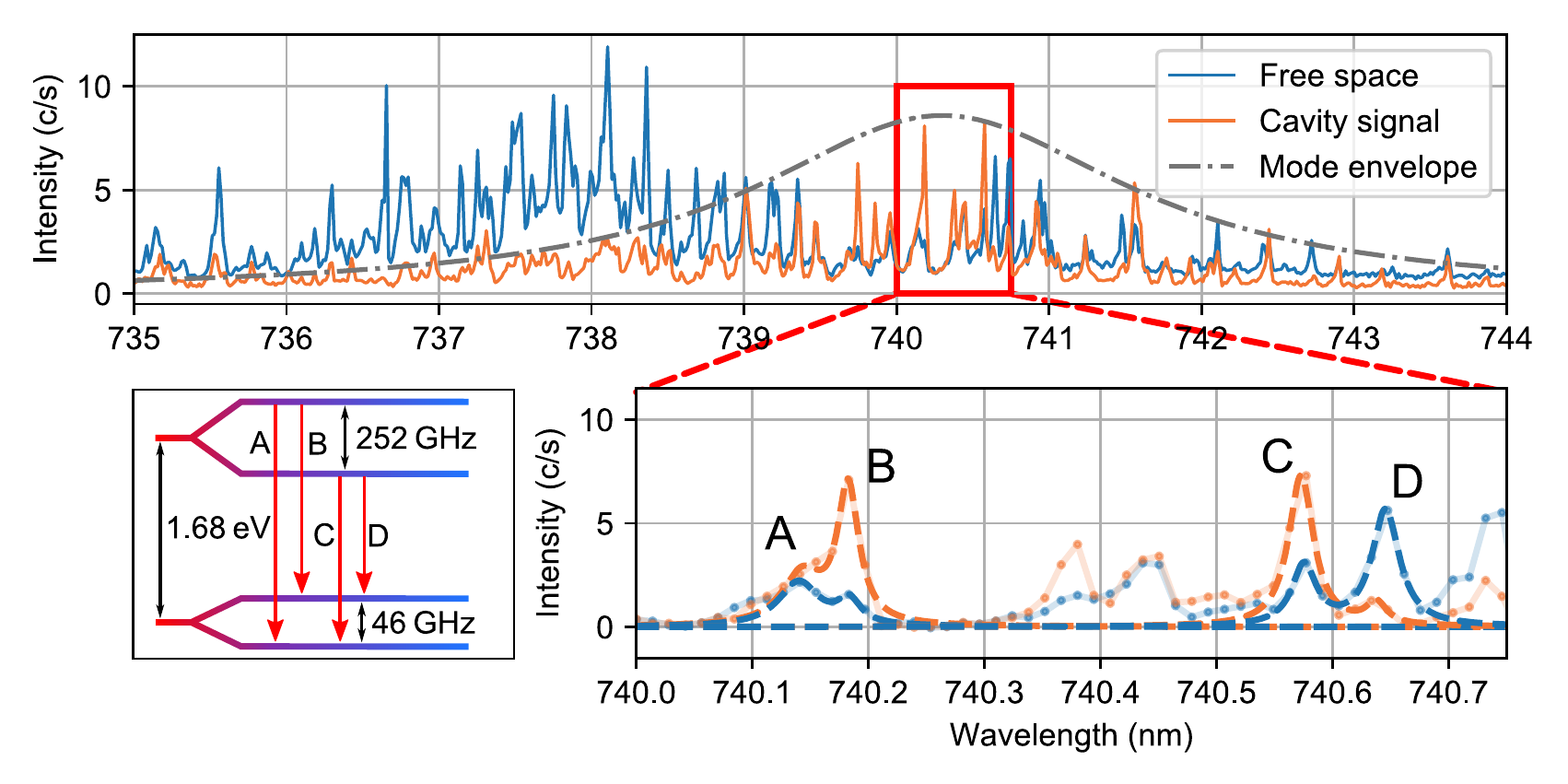}
\caption{Spectrum of coupled emitter-cavity system at cryogenic temperatures. Cavity emission of coupled \siv{} ensemble in orange. Free space emission in blue and envelope resonance mode in gray. For some spectral lines, the emission into the cavity mode reaches higher counting rates than into free space. The lower plot shows a zoom of the area in the red rectangle and the four transitions A-D (level scheme sketched on the right) of the \siv{} center are referenced to the lines according to their spacing.}
\label{fig:sivcoupling}
\end{figure*}

After the ND is positioned on top of the interaction zone, the sample is placed inside a flow-cryostat and cooled to approximately \SI{150}{\kelvin}. The emitter-cavity system is excited \textit{via} port 1 with \SI{130}{\micro\watt} of green laser while the emission is collected through port 3 (shown in Fig. \ref{fig:siv_warm} A in orange). In blue, the free space emission of the \siv{} center is shown, where the ensemble is evanescently excited through port 1 and the emission is collected at the center position of the ND. Both collection procedures are shown in the small insets in Fig. \ref{fig:siv_warm} A.\\
The presence of the ND in the interaction zone changes the effective refractive index of the cavity. This leads to a red shift of the desired cavity resonance. Together with a change in temperature the shift is approximately \SI{0.4}{\nano\meter}. The altered mode at $\sim \SI{737.9}{\nano\meter}$ of the PCC is fed by the ZPL of the \siv{} ensemble. The other resonances are slightly enhanced due to phonon side band coupling and scattered background fluorescence. The $Q$-factor of the desired resonance decreased from $Q\tief{cav}=1000$ for the empty cavity to $Q\tief{coupled}=480$ for the coupled system caused by scattering losses and  degradation of the PCC.\\
According to equation (\ref{eq:beta}), we obtain a coupling of:
\begin{equation}
\beta = \frac{\gammac/\eta_\mathrm{cav}}{\gammac/\eta_\mathrm{cav}+\gammaf/\eta_\mathrm{free}} = 0.44 \; ,
\end{equation}
where $\eta_\mathrm{cav}=0.14$ is the coupler efficiency and $\eta_\mathrm{free}=0.082$ is the collection efficiency of the free space emission. The coupler efficiency $\eta_\mathrm{cav}$ is determined \textit{via} a transmission measurement of a laser at the desired wavelength through the probe beam. The free space collection efficiency $\eta_\mathrm{free}$ is limited due to a NA=0.55 objective. Please note, that the dipole alignment of the \siv{} center is such that high free-space emission corresponds to the best coupled \siv{} centers, while our geometry does not allow for artificially decreased ratio between free-space and waveguide emission. The achieved $\beta$-factor corresponds to nearly every second photon being emitted into the cavity resonance and leads to an average Purcell-enhancement
\cite{zhang_strongly_2018} of 
\begin{equation}
F\tief{P} = \frac{\beta}{1-\beta} \approx 0.79
\end{equation}
for all emitters inside the \siv{} ensemble. 
The tuning mechanisms for modifying the ensemble coupling utilized so far, are position of the ND and temperature. In Fig. \ref{fig:siv_warm} B the $\beta$-factors after optimizing the position of the ND and after temperature tuning are shown. The position and orientation of the ND in the interaction zone can be controlled by AFM-based nanomanipulation
\cite{hausler_preparing_2019},
which yields to a better dipole alignment of the ND to the cavity axis. Position 1 and position 2 in Fig. \ref{fig:siv_warm} B correspond to a coupling averaged over the whole ensemble of 0.14 and 0.31, respectively. At position 2 we cooled the sample from \SI{295}{\kelvin} to $\sim \SI{150}{\kelvin}$. This further increased the average ensemble $\beta$-factor to 0.44 (as shown in Fig. \ref{fig:siv_warm} A and given in equation (\ref{eq:beta})).\\
Additional reduction of the temperature to $\sim$\SI{4}{\kelvin} reveals the fine structure splitting of the \siv{} centers in the ensemble. Thus, instead of the average ensemble coupling the $\beta$-factors of individual transitions of single \siv{} centers can be determined, as depicted in Fig. \ref{fig:sivcoupling}.
Again, the free space emission (blue) needs to be compared to the joint emitter cavity signal (insets Fig. \ref{fig:siv_warm} A). Freezing and unfreezing processes, together with temperature tuning further shifted the central frequency of the resonance mode to approximately \SI{740.3}{\nano\meter} (depicted in gray Fig. \ref{fig:sivcoupling} top). This implies a detuning from the average ensemble ZPL resonance at $\sim \SI{738}{\nano\meter}$ leading to decreased average coupling. The detuning enables the coupling of individual, however more strained, \siv{} centers (Fig. \ref{fig:sivcoupling} top), apparent when zooming into the red square (Fig. \ref{fig:sivcoupling} bottom). A splitting of \SI[separate-uncertainty = true]{244(10)}{\giga\hertz} between the doublet of (A,B) and (C,D) is observed, while the splitting between C and D is \SI[separate-uncertainty = true]{37(10)}{\giga\hertz}. These four lines arise from one \siv{} center with an excited state splitting of \SI{252}{\giga\hertz} and a ground state splitting of \SI{46.3}{\giga\hertz} influenced by the strain inside the ND \cite{rogers_single_2019}, where the ZPL position of the \siv{} can be shifted (axial strain) while keeping the state splitting constant (low transverse strain)\cite{meesala_strain_2018}. The used NDs have proven to inhere low transverse strain \cite{rogers_single_2019}. For the \siv{} center transition A and D have the same dipole orientation as well as transition B and C \cite{rogers_all-optical_2014}. Fig. \ref{fig:sivcoupling} confirms that transition B and C show a better coupling to the cavity mode than transition A and D originating from dipole alignment of the emitter to the cavity axis. Due to strain induced differences in the selection rules, transition B and C show different enhancement factors \cite{hepp_electronic_2014}. 

 Similar dipole orientation of the transitions reach higher (B and C) and lower (A and D) coupling efficiencies to the cavity mode. From these values according to equation (\ref{eq:beta}), we estimate the $\beta$-factor for all four transitions. The highest coupling was achieved for transition B, which is $\beta\tief{B} = 0.81$. This value translates to a Purcell enhancement of 
\[F\tief{P,B} \approx 4.\]
4/5 of the total spontaneous emission are channeled into the photonic circuit. The measured $\beta$-factor of $0.81$ corresponds to a lifetime reduction of $0.2 \tau _0$, where $\tau _0$ is the lifetime of the \siv{} without cavity. The resulting lifetime for the \siv{} on resonance with the cavity is about $340$ ps, too short to be observed with our current experimental equipment.
\section*{Outlook}
In our work we demonstrate the efficient coupling of individual optical transitions of \siv{} centers to a \sn{}-based cross-bar PCC with a Purcell enhancement of more than 4, where the enhanced signal was read out \textit{via} a probe waveguide (with a PCC) by excitation \textit{via} a crossed waveguide on chip. Much higher Purcell enhancement is prohibited by residual scattering from the ND reducing the overall Q-factor of the PCC. 
 Furthermore, the achievable evanescent coupling strength is ultimately limited by the distance of the \siv{} center to the field maximum of the PCC. In future experiments we envision Purcell factors beyond 100 for single \siv{} center in nanodiamonds with diameter of a few ten nanometers.
To emphasize the influence of the ND emitter host on the Purcell enhancement factor simulations without the scattering crystal were computed leading to $F\tief{P}=68$ ($\beta=0.98$; $N=27$) and $F\tief{P}=541$ ($\beta=0.998$; $N=53$), respectively for matched polarization of the emitter-cavity systems. 
For the first time, we experimentally demonstrate Purcell enhancement well above one for hybrid quantum photonics with SiV center in NDs matching recently simulated Purcell factors of $\sim 8$ \cite{radulaski_nanodiamond_2019}. In agreement with our simulations, Purcell factors of more than 500 were envisioned for less-degraded systems \cite{radulaski_nanodiamond_2019}.
The investigated hybrid quantum photonics platform brings cavity-mediated entanglement generation \cite{imamoglu_quantum_1999,kastoryano_dissipative_2011,zheng_efficient_2000},
efficient Bell-State measurements 
\cite{waks_dipole_2006, borregaard_long-distance_2015}
and robust gates of distant emitters 
\cite{cirac_quantum_1997}
into reach. Quantum state transfer in an on-chip, integrated platform opens the door for long-distance quantum communication and linear optics quantum computing
\cite{kok_publishers_2007}. Quantum photonics based on \siv{} center in diamond, where the electronic spin is coupled to the environment whereas the nuclear spin is well-isolated, enable further applications relying on the electronic spin as broker unit with a connected, long-lived quantum memory. Applications range from photonic memories 
\cite{de_riedmatten_quantum_2015}
  and quantum repeaters 
\cite{munro_inside_2015}
  to error correction 
\cite{terhal_quantum_2015}
   or enhanced quantum sensing
\cite{degen_quantum_2017}.

\section*{Methods}
\subsection*{Fabrication of Nanodiamonds}
\siv{} containing  diamond nanoparticles were obtained by HPHT treatment of the metal catalysts-free growth system based on a homogeneous mixture of naphthalene (C$_{10}$H$_8$), fluorographite (CF$_{1.1}$), and tetrakis(trimethylsilyl)silane (C$_{12}$H$_{36}$Si$_5$) which was used as the Si doping component. (Introduction of fluorine-containing compounds into the growth system was intended to reduce the content of NV centers in NDs-\siv{})
HPHT treatment of the initial homogeneous mixtures was carried out in a high-pressure apparatus of "Toroid" type. The experimental procedure consists of loading the high-pressure apparatus to \SI{8.0}{\giga\pascal}, heating the samples up to \SI{1450}{\celsius} and short (\SI{3}{\second}) isothermal exposures at these temperatures.

\subsection*{Fabrication of the Photonic Chip}
Free-standing PCC devices on chip were realized on Silicon nitride-on-insulator wafers consisting of \SI{200}{\nano\meter} stoichiometric \sn{} on top of a \SI{2}{\micro\meter} thick SiO$_2$ layer on top of Si. Fabrication of the nanophotonic circuits involved several steps of electron-beam lithography (e-beam) followed by reactive ion etching. The nanophotonic circuits were defined on top of the \sn{} layer using negative tone photoresist ma-N 2403 in the first lithography step and after were \SI{75}{\percent} dry-etched into the silicon nitride layer using an  CHF$_3$/O$_2$ plasma. To realize freestanding PCC underneath SiO$_2$ should be removed, which was achieved by opening a window around the photonic crystal region in the second step of lithography by means of exposing
positive photoresist PMMA in this area. In the following step, the remaining \SI{25}{\percent} of silicon nitride in the window area was etched, while the waveguide inside of the window was protected with a ma-N 2403 photoresist; the waveguides outside the windows were protected by unexposed PMMA photoresist. After that, both photoresists were removed by O$_2$ plasma. In the last fabrication step SiO$_2$ layer in the windows was removed by wet etching, namely by immersing the chip in hydrofluoric acid (HF).

\subsection*{Optical Methods}
The optical readout was established by a self build confocal microscope setup with an NA=0.55 objective (50x magnification). For scanning the sample, a galvo mirror system was used. With the help of a second laser path and a 4f-lens system, the laser can be set on a fixed position, while the readout is collected from a different position. The readout can be directed on a spectrometer and an avalanche photo diode.
%\the\columnwidth
%\printinunitsof{cm}\prntlen{\textwidth}

\section*{Acknowledgment}
The project was funded by the Deutsche Forschungsgemeinschaft (DFG, German
Research Foundation) - Project number: 398628099. AK acknowledges support of
the BMBF/VDI in project Q.Link.X and the European fund for regional development (EFRE) program Baden-Württemberg. KGF and AK acknowledge support of IQst. The AFM was funded by the DFG. We thank Prof. Kay Gottschalk and Frederike Erb for their support. V. A. Davydov thanks the Russian Foundation for Basic Research (Grant No. 18-03-00936) for financial support.


\begin{thebibliography}{40}%
\makeatletter
\providecommand \@ifxundefined [1]{%
 \@ifx{#1\undefined}
}%
\providecommand \@ifnum [1]{%
 \ifnum #1\expandafter \@firstoftwo
 \else \expandafter \@secondoftwo
 \fi
}%
\providecommand \@ifx [1]{%
 \ifx #1\expandafter \@firstoftwo
 \else \expandafter \@secondoftwo
 \fi
}%
\providecommand \natexlab [1]{#1}%
\providecommand \enquote  [1]{``#1''}%
\providecommand \bibnamefont  [1]{#1}%
\providecommand \bibfnamefont [1]{#1}%
\providecommand \citenamefont [1]{#1}%
\providecommand \href@noop [0]{\@secondoftwo}%
\providecommand \href [0]{\begingroup \@sanitize@url \@href}%
\providecommand \@href[1]{\@@startlink{#1}\@@href}%
\providecommand \@@href[1]{\endgroup#1\@@endlink}%
\providecommand \@sanitize@url [0]{\catcode `\\12\catcode `\$12\catcode
  `\&12\catcode `\#12\catcode `\^12\catcode `\_12\catcode `\%12\relax}%
\providecommand \@@startlink[1]{}%
\providecommand \@@endlink[0]{}%
\providecommand \url  [0]{\begingroup\@sanitize@url \@url }%
\providecommand \@url [1]{\endgroup\@href {#1}{\urlprefix }}%
\providecommand \urlprefix  [0]{URL }%
\providecommand \Eprint [0]{\href }%
\providecommand \doibase [0]{http://dx.doi.org/}%
\providecommand \selectlanguage [0]{\@gobble}%
\providecommand \bibinfo  [0]{\@secondoftwo}%
\providecommand \bibfield  [0]{\@secondoftwo}%
\providecommand \translation [1]{[#1]}%
\providecommand \BibitemOpen [0]{}%
\providecommand \bibitemStop [0]{}%
\providecommand \bibitemNoStop [0]{.\EOS\space}%
\providecommand \EOS [0]{\spacefactor3000\relax}%
\providecommand \BibitemShut  [1]{\csname bibitem#1\endcsname}%
\let\auto@bib@innerbib\@empty
%</preamble>
\bibitem [{\citenamefont {Atatüre}\ \emph {et~al.}(2018)\citenamefont
  {Atatüre}, \citenamefont {Englund}, \citenamefont {Vamivakas}, \citenamefont
  {Lee},\ and\ \citenamefont {Wrachtrup}}]{atature_material_2018}%
  \BibitemOpen
  \bibfield  {author} {\bibinfo {author} {\bibfnamefont {M.}~\bibnamefont
  {Atatüre}}, \bibinfo {author} {\bibfnamefont {D.}~\bibnamefont {Englund}},
  \bibinfo {author} {\bibfnamefont {N.}~\bibnamefont {Vamivakas}}, \bibinfo
  {author} {\bibfnamefont {S.-Y.}\ \bibnamefont {Lee}}, \ and\ \bibinfo
  {author} {\bibfnamefont {J.}~\bibnamefont {Wrachtrup}},\ }\href {\doibase
  10.1038/s41578-018-0008-9} {\bibfield  {journal} {\bibinfo  {journal} {Nat.
  Rev. Mater.}\ }\textbf {\bibinfo {volume} {3}},\ \bibinfo {pages} {38}
  (\bibinfo {year} {2018})}\BibitemShut {NoStop}%
\bibitem [{\citenamefont {Awschalom}\ \emph {et~al.}(2018)\citenamefont
  {Awschalom}, \citenamefont {Hanson}, \citenamefont {Wrachtrup},\ and\
  \citenamefont {Zhou}}]{awschalom_quantum_2018}%
  \BibitemOpen
  \bibfield  {author} {\bibinfo {author} {\bibfnamefont {D.~D.}\ \bibnamefont
  {Awschalom}}, \bibinfo {author} {\bibfnamefont {R.}~\bibnamefont {Hanson}},
  \bibinfo {author} {\bibfnamefont {J.}~\bibnamefont {Wrachtrup}}, \ and\
  \bibinfo {author} {\bibfnamefont {B.~B.}\ \bibnamefont {Zhou}},\ }\href
  {\doibase 10.1038/s41566-018-0232-2} {\bibfield  {journal} {\bibinfo
  {journal} {Nat. Photonics}\ }\textbf {\bibinfo {volume} {12}},\ \bibinfo
  {pages} {516} (\bibinfo {year} {2018})}\BibitemShut {NoStop}%
\bibitem [{\citenamefont {Becker}\ and\ \citenamefont
  {Becher}(2017)}]{becker_coherence_2017}%
  \BibitemOpen
  \bibfield  {author} {\bibinfo {author} {\bibfnamefont {J.~N.}\ \bibnamefont
  {Becker}}\ and\ \bibinfo {author} {\bibfnamefont {C.}~\bibnamefont
  {Becher}},\ }\href@noop {} {\bibfield  {journal} {\bibinfo  {journal} {Phys.
  Status Solidi A}\ }\textbf {\bibinfo {volume} {214}},\ \bibinfo {pages}
  {1700586} (\bibinfo {year} {2017})}\BibitemShut {NoStop}%
\bibitem [{\citenamefont {Rogers}\ \emph {et~al.}(2019)\citenamefont {Rogers},
  \citenamefont {Wang}, \citenamefont {Liu}, \citenamefont {Antoniuk},
  \citenamefont {Osterkamp}, \citenamefont {Davydov}, \citenamefont {Agafonov},
  \citenamefont {Filipovski}, \citenamefont {Jelezko},\ and\ \citenamefont
  {Kubanek}}]{rogers_single_2019}%
  \BibitemOpen
  \bibfield  {author} {\bibinfo {author} {\bibfnamefont {L.~J.}\ \bibnamefont
  {Rogers}}, \bibinfo {author} {\bibfnamefont {O.}~\bibnamefont {Wang}},
  \bibinfo {author} {\bibfnamefont {Y.}~\bibnamefont {Liu}}, \bibinfo {author}
  {\bibfnamefont {L.}~\bibnamefont {Antoniuk}}, \bibinfo {author}
  {\bibfnamefont {C.}~\bibnamefont {Osterkamp}}, \bibinfo {author}
  {\bibfnamefont {V.~A.}\ \bibnamefont {Davydov}}, \bibinfo {author}
  {\bibfnamefont {V.~N.}\ \bibnamefont {Agafonov}}, \bibinfo {author}
  {\bibfnamefont {A.~B.}\ \bibnamefont {Filipovski}}, \bibinfo {author}
  {\bibfnamefont {F.}~\bibnamefont {Jelezko}}, \ and\ \bibinfo {author}
  {\bibfnamefont {A.}~\bibnamefont {Kubanek}},\ }\href {\doibase
  10.1103/PhysRevApplied.11.024073} {\bibfield  {journal} {\bibinfo  {journal}
  {Phys. Rev. Appl.}\ }\textbf {\bibinfo {volume} {11}},\ \bibinfo {pages}
  {024073} (\bibinfo {year} {2019})}\BibitemShut {NoStop}%
\bibitem [{\citenamefont {Sipahigil}\ \emph {et~al.}(2014)\citenamefont
  {Sipahigil}, \citenamefont {Jahnke}, \citenamefont {Rogers}, \citenamefont
  {Teraji}, \citenamefont {Isoya}, \citenamefont {Zibrov}, \citenamefont
  {Jelezko},\ and\ \citenamefont {Lukin}}]{sipahigil_indistinguishable_2014}%
  \BibitemOpen
  \bibfield  {author} {\bibinfo {author} {\bibfnamefont {A.}~\bibnamefont
  {Sipahigil}}, \bibinfo {author} {\bibfnamefont {K.~D.}\ \bibnamefont
  {Jahnke}}, \bibinfo {author} {\bibfnamefont {L.~J.}\ \bibnamefont {Rogers}},
  \bibinfo {author} {\bibfnamefont {T.}~\bibnamefont {Teraji}}, \bibinfo
  {author} {\bibfnamefont {J.}~\bibnamefont {Isoya}}, \bibinfo {author}
  {\bibfnamefont {A.~S.}\ \bibnamefont {Zibrov}}, \bibinfo {author}
  {\bibfnamefont {F.}~\bibnamefont {Jelezko}}, \ and\ \bibinfo {author}
  {\bibfnamefont {M.~D.}\ \bibnamefont {Lukin}},\ }\href {\doibase
  10.1103/PhysRevLett.113.113602} {\bibfield  {journal} {\bibinfo  {journal}
  {Phys. Rev. Lett.}\ }\textbf {\bibinfo {volume} {113}},\ \bibinfo {pages}
  {113602} (\bibinfo {year} {2014})}\BibitemShut {NoStop}%
\bibitem [{\citenamefont {Sohn}\ \emph {et~al.}(2018)\citenamefont {Sohn},
  \citenamefont {Meesala}, \citenamefont {Pingault}, \citenamefont {Atikian},
  \citenamefont {Holzgrafe}, \citenamefont {G{\"u}ndo{\u{g}}an}, \citenamefont
  {Stavrakas}, \citenamefont {Stanley}, \citenamefont {Sipahigil},
  \citenamefont {Choi} \emph {et~al.}}]{sohn_controlling_2018}%
  \BibitemOpen
  \bibfield  {author} {\bibinfo {author} {\bibfnamefont {Y.-I.}\ \bibnamefont
  {Sohn}}, \bibinfo {author} {\bibfnamefont {S.}~\bibnamefont {Meesala}},
  \bibinfo {author} {\bibfnamefont {B.}~\bibnamefont {Pingault}}, \bibinfo
  {author} {\bibfnamefont {H.~A.}\ \bibnamefont {Atikian}}, \bibinfo {author}
  {\bibfnamefont {J.}~\bibnamefont {Holzgrafe}}, \bibinfo {author}
  {\bibfnamefont {M.}~\bibnamefont {G{\"u}ndo{\u{g}}an}}, \bibinfo {author}
  {\bibfnamefont {C.}~\bibnamefont {Stavrakas}}, \bibinfo {author}
  {\bibfnamefont {M.~J.}\ \bibnamefont {Stanley}}, \bibinfo {author}
  {\bibfnamefont {A.}~\bibnamefont {Sipahigil}}, \bibinfo {author}
  {\bibfnamefont {J.}~\bibnamefont {Choi}},  \emph {et~al.},\ }\href@noop {}
  {\bibfield  {journal} {\bibinfo  {journal} {Nat. Commun.}\ }\textbf {\bibinfo
  {volume} {9}},\ \bibinfo {pages} {1} (\bibinfo {year} {2018})}\BibitemShut
  {NoStop}%
\bibitem [{\citenamefont {Sukachev}\ \emph {et~al.}(2017)\citenamefont
  {Sukachev}, \citenamefont {Sipahigil}, \citenamefont {Nguyen}, \citenamefont
  {Bhaskar}, \citenamefont {Evans}, \citenamefont {Jelezko},\ and\
  \citenamefont {Lukin}}]{sukachev_silicon_2017}%
  \BibitemOpen
  \bibfield  {author} {\bibinfo {author} {\bibfnamefont {D.~D.}\ \bibnamefont
  {Sukachev}}, \bibinfo {author} {\bibfnamefont {A.}~\bibnamefont {Sipahigil}},
  \bibinfo {author} {\bibfnamefont {C.~T.}\ \bibnamefont {Nguyen}}, \bibinfo
  {author} {\bibfnamefont {M.~K.}\ \bibnamefont {Bhaskar}}, \bibinfo {author}
  {\bibfnamefont {R.~E.}\ \bibnamefont {Evans}}, \bibinfo {author}
  {\bibfnamefont {F.}~\bibnamefont {Jelezko}}, \ and\ \bibinfo {author}
  {\bibfnamefont {M.~D.}\ \bibnamefont {Lukin}},\ }\href@noop {} {\bibfield
  {journal} {\bibinfo  {journal} {Phys. Rev. Lett.}\ }\textbf {\bibinfo
  {volume} {119}},\ \bibinfo {pages} {223602} (\bibinfo {year}
  {2017})}\BibitemShut {NoStop}%
\bibitem [{\citenamefont {Metsch}\ \emph {et~al.}(2019)\citenamefont {Metsch},
  \citenamefont {Senkalla}, \citenamefont {Tratzmiller}, \citenamefont
  {Scheuer}, \citenamefont {Kern}, \citenamefont {Achard}, \citenamefont
  {Tallaire}, \citenamefont {Plenio}, \citenamefont {Siyushev},\ and\
  \citenamefont {Jelezko}}]{metsch_initialization_2019}%
  \BibitemOpen
  \bibfield  {author} {\bibinfo {author} {\bibfnamefont {M.~H.}\ \bibnamefont
  {Metsch}}, \bibinfo {author} {\bibfnamefont {K.}~\bibnamefont {Senkalla}},
  \bibinfo {author} {\bibfnamefont {B.}~\bibnamefont {Tratzmiller}}, \bibinfo
  {author} {\bibfnamefont {J.}~\bibnamefont {Scheuer}}, \bibinfo {author}
  {\bibfnamefont {M.}~\bibnamefont {Kern}}, \bibinfo {author} {\bibfnamefont
  {J.}~\bibnamefont {Achard}}, \bibinfo {author} {\bibfnamefont
  {A.}~\bibnamefont {Tallaire}}, \bibinfo {author} {\bibfnamefont {M.~B.}\
  \bibnamefont {Plenio}}, \bibinfo {author} {\bibfnamefont {P.}~\bibnamefont
  {Siyushev}}, \ and\ \bibinfo {author} {\bibfnamefont {F.}~\bibnamefont
  {Jelezko}},\ }\href@noop {} {\bibfield  {journal} {\bibinfo  {journal} {Phys.
  Rev. Lett.}\ }\textbf {\bibinfo {volume} {122}},\ \bibinfo {pages} {190503}
  (\bibinfo {year} {2019})}\BibitemShut {NoStop}%
\bibitem [{\citenamefont {Fu}\ \emph {et~al.}(2008)\citenamefont {Fu},
  \citenamefont {Santori}, \citenamefont {Barclay}, \citenamefont
  {Aharonovich}, \citenamefont {Prawer}, \citenamefont {Meyer}, \citenamefont
  {Holm},\ and\ \citenamefont {Beausoleil}}]{fu_coupling_2008}%
  \BibitemOpen
  \bibfield  {author} {\bibinfo {author} {\bibfnamefont {K.-M.~C.}\
  \bibnamefont {Fu}}, \bibinfo {author} {\bibfnamefont {C.}~\bibnamefont
  {Santori}}, \bibinfo {author} {\bibfnamefont {P.~E.}\ \bibnamefont
  {Barclay}}, \bibinfo {author} {\bibfnamefont {I.}~\bibnamefont
  {Aharonovich}}, \bibinfo {author} {\bibfnamefont {S.}~\bibnamefont {Prawer}},
  \bibinfo {author} {\bibfnamefont {N.}~\bibnamefont {Meyer}}, \bibinfo
  {author} {\bibfnamefont {A.~M.}\ \bibnamefont {Holm}}, \ and\ \bibinfo
  {author} {\bibfnamefont {R.~G.}\ \bibnamefont {Beausoleil}},\ }\href
  {\doibase 10.1063/1.3045950} {\bibfield  {journal} {\bibinfo  {journal}
  {Appl. Phys. Lett.}\ }\textbf {\bibinfo {volume} {93}},\ \bibinfo {pages}
  {234107} (\bibinfo {year} {2008})}\BibitemShut {NoStop}%
\bibitem [{\citenamefont {Englund}\ \emph {et~al.}(2010)\citenamefont
  {Englund}, \citenamefont {Shields}, \citenamefont {Rivoire}, \citenamefont
  {Hatami}, \citenamefont {Vučković}, \citenamefont {Park},\ and\
  \citenamefont {Lukin}}]{englund_deterministic_2010}%
  \BibitemOpen
  \bibfield  {author} {\bibinfo {author} {\bibfnamefont {D.}~\bibnamefont
  {Englund}}, \bibinfo {author} {\bibfnamefont {B.}~\bibnamefont {Shields}},
  \bibinfo {author} {\bibfnamefont {K.}~\bibnamefont {Rivoire}}, \bibinfo
  {author} {\bibfnamefont {F.}~\bibnamefont {Hatami}}, \bibinfo {author}
  {\bibfnamefont {J.}~\bibnamefont {Vučković}}, \bibinfo {author}
  {\bibfnamefont {H.}~\bibnamefont {Park}}, \ and\ \bibinfo {author}
  {\bibfnamefont {M.~D.}\ \bibnamefont {Lukin}},\ }\href {\doibase
  10.1021/nl101662v} {\bibfield  {journal} {\bibinfo  {journal} {Nano Lett.}\
  }\textbf {\bibinfo {volume} {10}},\ \bibinfo {pages} {3922} (\bibinfo {year}
  {2010})}\BibitemShut {NoStop}%
\bibitem [{\citenamefont {Barclay}\ \emph {et~al.}(2011)\citenamefont
  {Barclay}, \citenamefont {Fu}, \citenamefont {Santori}, \citenamefont
  {Faraon},\ and\ \citenamefont {Beausoleil}}]{barclay_hybrid_2011}%
  \BibitemOpen
  \bibfield  {author} {\bibinfo {author} {\bibfnamefont {P.~E.}\ \bibnamefont
  {Barclay}}, \bibinfo {author} {\bibfnamefont {K.-M.~C.}\ \bibnamefont {Fu}},
  \bibinfo {author} {\bibfnamefont {C.}~\bibnamefont {Santori}}, \bibinfo
  {author} {\bibfnamefont {A.}~\bibnamefont {Faraon}}, \ and\ \bibinfo {author}
  {\bibfnamefont {R.~G.}\ \bibnamefont {Beausoleil}},\ }\href {\doibase
  10.1103/PhysRevX.1.011007} {\bibfield  {journal} {\bibinfo  {journal} {Phys.
  Rev. X}\ }\textbf {\bibinfo {volume} {1}},\ \bibinfo {pages} {011007}
  (\bibinfo {year} {2011})}\BibitemShut {NoStop}%
\bibitem [{\citenamefont {Wolters}\ \emph {et~al.}(2010)\citenamefont
  {Wolters}, \citenamefont {Schell}, \citenamefont {Kewes}, \citenamefont
  {Nüsse}, \citenamefont {Schoengen}, \citenamefont {Döscher}, \citenamefont
  {Hannappel}, \citenamefont {Löchel}, \citenamefont {Barth},\ and\
  \citenamefont {Benson}}]{wolters_enhancement_2010}%
  \BibitemOpen
  \bibfield  {author} {\bibinfo {author} {\bibfnamefont {J.}~\bibnamefont
  {Wolters}}, \bibinfo {author} {\bibfnamefont {A.~W.}\ \bibnamefont {Schell}},
  \bibinfo {author} {\bibfnamefont {G.}~\bibnamefont {Kewes}}, \bibinfo
  {author} {\bibfnamefont {N.}~\bibnamefont {Nüsse}}, \bibinfo {author}
  {\bibfnamefont {M.}~\bibnamefont {Schoengen}}, \bibinfo {author}
  {\bibfnamefont {H.}~\bibnamefont {Döscher}}, \bibinfo {author}
  {\bibfnamefont {T.}~\bibnamefont {Hannappel}}, \bibinfo {author}
  {\bibfnamefont {B.}~\bibnamefont {Löchel}}, \bibinfo {author} {\bibfnamefont
  {M.}~\bibnamefont {Barth}}, \ and\ \bibinfo {author} {\bibfnamefont
  {O.}~\bibnamefont {Benson}},\ }\href {\doibase 10.1063/1.3499300} {\bibfield
  {journal} {\bibinfo  {journal} {Appl. Phys. Lett.}\ }\textbf {\bibinfo
  {volume} {97}},\ \bibinfo {pages} {141108} (\bibinfo {year}
  {2010})}\BibitemShut {NoStop}%
\bibitem [{\citenamefont {Radulaski}\ \emph {et~al.}(2019)\citenamefont
  {Radulaski}, \citenamefont {Zhang}, \citenamefont {Tzeng}, \citenamefont
  {Lagoudakis}, \citenamefont {Ishiwata}, \citenamefont {Dory}, \citenamefont
  {Fischer}, \citenamefont {Kelaita}, \citenamefont {Sun}, \citenamefont
  {Maurer} \emph {et~al.}}]{radulaski_nanodiamond_2019}%
  \BibitemOpen
  \bibfield  {author} {\bibinfo {author} {\bibfnamefont {M.}~\bibnamefont
  {Radulaski}}, \bibinfo {author} {\bibfnamefont {J.~L.}\ \bibnamefont
  {Zhang}}, \bibinfo {author} {\bibfnamefont {Y.-K.}\ \bibnamefont {Tzeng}},
  \bibinfo {author} {\bibfnamefont {K.~G.}\ \bibnamefont {Lagoudakis}},
  \bibinfo {author} {\bibfnamefont {H.}~\bibnamefont {Ishiwata}}, \bibinfo
  {author} {\bibfnamefont {C.}~\bibnamefont {Dory}}, \bibinfo {author}
  {\bibfnamefont {K.~A.}\ \bibnamefont {Fischer}}, \bibinfo {author}
  {\bibfnamefont {Y.~A.}\ \bibnamefont {Kelaita}}, \bibinfo {author}
  {\bibfnamefont {S.}~\bibnamefont {Sun}}, \bibinfo {author} {\bibfnamefont
  {P.~C.}\ \bibnamefont {Maurer}},  \emph {et~al.},\ }\href@noop {} {\bibfield
  {journal} {\bibinfo  {journal} {Laser Photonics Rev.}\ }\textbf {\bibinfo
  {volume} {13}},\ \bibinfo {pages} {1800316} (\bibinfo {year}
  {2019})}\BibitemShut {NoStop}%
\bibitem [{\citenamefont {Fehler}\ \emph {et~al.}(2019)\citenamefont {Fehler},
  \citenamefont {Ovvyan}, \citenamefont {Gruhler}, \citenamefont {Pernice},\
  and\ \citenamefont {Kubanek}}]{fehler_efficient_2019}%
  \BibitemOpen
  \bibfield  {author} {\bibinfo {author} {\bibfnamefont {K.~G.}\ \bibnamefont
  {Fehler}}, \bibinfo {author} {\bibfnamefont {A.~P.}\ \bibnamefont {Ovvyan}},
  \bibinfo {author} {\bibfnamefont {N.}~\bibnamefont {Gruhler}}, \bibinfo
  {author} {\bibfnamefont {W.~H.~P.}\ \bibnamefont {Pernice}}, \ and\ \bibinfo
  {author} {\bibfnamefont {A.}~\bibnamefont {Kubanek}},\ }\href {\doibase
  10.1021/acsnano.9b01668} {\bibfield  {journal} {\bibinfo  {journal} {ACS
  Nano}\ }\textbf {\bibinfo {volume} {13}},\ \bibinfo {pages} {6891} (\bibinfo
  {year} {2019})}\BibitemShut {NoStop}%
\bibitem [{\citenamefont {Jantzen}\ \emph {et~al.}(2016)\citenamefont
  {Jantzen}, \citenamefont {Kurz}, \citenamefont {Rudnicki}, \citenamefont
  {Schäfermeier}, \citenamefont {Jahnke}, \citenamefont {Andersen},
  \citenamefont {Davydov}, \citenamefont {Agafonov}, \citenamefont {Kubanek},
  \citenamefont {Rogers},\ and\ \citenamefont
  {Jelezko}}]{jantzen_nanodiamonds_2016}%
  \BibitemOpen
  \bibfield  {author} {\bibinfo {author} {\bibfnamefont {U.}~\bibnamefont
  {Jantzen}}, \bibinfo {author} {\bibfnamefont {A.~B.}\ \bibnamefont {Kurz}},
  \bibinfo {author} {\bibfnamefont {D.~S.}\ \bibnamefont {Rudnicki}}, \bibinfo
  {author} {\bibfnamefont {C.}~\bibnamefont {Schäfermeier}}, \bibinfo {author}
  {\bibfnamefont {K.~D.}\ \bibnamefont {Jahnke}}, \bibinfo {author}
  {\bibfnamefont {U.~L.}\ \bibnamefont {Andersen}}, \bibinfo {author}
  {\bibfnamefont {V.~A.}\ \bibnamefont {Davydov}}, \bibinfo {author}
  {\bibfnamefont {V.~N.}\ \bibnamefont {Agafonov}}, \bibinfo {author}
  {\bibfnamefont {A.}~\bibnamefont {Kubanek}}, \bibinfo {author} {\bibfnamefont
  {L.~J.}\ \bibnamefont {Rogers}}, \ and\ \bibinfo {author} {\bibfnamefont
  {F.}~\bibnamefont {Jelezko}},\ }\href {\doibase
  10.1088/1367-2630/18/7/073036} {\bibfield  {journal} {\bibinfo  {journal}
  {New J. Phys.}\ }\textbf {\bibinfo {volume} {18}},\ \bibinfo {pages} {073036}
  (\bibinfo {year} {2016})}\BibitemShut {NoStop}%
\bibitem [{\citenamefont {Häußler}\ \emph {et~al.}(2019)\citenamefont
  {Häußler}, \citenamefont {Hartung}, \citenamefont {Fehler}, \citenamefont
  {Antoniuk}, \citenamefont {Kulikova}, \citenamefont {Davydov}, \citenamefont
  {Agafonov}, \citenamefont {Jelezko},\ and\ \citenamefont
  {Kubanek}}]{hausler_preparing_2019}%
  \BibitemOpen
  \bibfield  {author} {\bibinfo {author} {\bibfnamefont {S.}~\bibnamefont
  {Häußler}}, \bibinfo {author} {\bibfnamefont {L.}~\bibnamefont {Hartung}},
  \bibinfo {author} {\bibfnamefont {K.~G.}\ \bibnamefont {Fehler}}, \bibinfo
  {author} {\bibfnamefont {L.}~\bibnamefont {Antoniuk}}, \bibinfo {author}
  {\bibfnamefont {L.~F.}\ \bibnamefont {Kulikova}}, \bibinfo {author}
  {\bibfnamefont {V.~A.}\ \bibnamefont {Davydov}}, \bibinfo {author}
  {\bibfnamefont {V.~N.}\ \bibnamefont {Agafonov}}, \bibinfo {author}
  {\bibfnamefont {F.}~\bibnamefont {Jelezko}}, \ and\ \bibinfo {author}
  {\bibfnamefont {A.}~\bibnamefont {Kubanek}},\ }\href@noop {} {\bibfield
  {journal} {\bibinfo  {journal} {New J. Phys.}\ }\textbf {\bibinfo {volume}
  {21}},\ \bibinfo {pages} {103047} (\bibinfo {year} {2019})}\BibitemShut
  {NoStop}%
\bibitem [{\citenamefont {Schell}\ \emph {et~al.}(2011)\citenamefont {Schell},
  \citenamefont {Kewes}, \citenamefont {Schröder}, \citenamefont {Wolters},
  \citenamefont {Aichele},\ and\ \citenamefont
  {Benson}}]{schell_scanning_2011}%
  \BibitemOpen
  \bibfield  {author} {\bibinfo {author} {\bibfnamefont {A.~W.}\ \bibnamefont
  {Schell}}, \bibinfo {author} {\bibfnamefont {G.}~\bibnamefont {Kewes}},
  \bibinfo {author} {\bibfnamefont {T.}~\bibnamefont {Schröder}}, \bibinfo
  {author} {\bibfnamefont {J.}~\bibnamefont {Wolters}}, \bibinfo {author}
  {\bibfnamefont {T.}~\bibnamefont {Aichele}}, \ and\ \bibinfo {author}
  {\bibfnamefont {O.}~\bibnamefont {Benson}},\ }\href {\doibase
  10.1063/1.3615629} {\bibfield  {journal} {\bibinfo  {journal} {Rev. Sci.
  Instrum.}\ }\textbf {\bibinfo {volume} {82}},\ \bibinfo {pages} {073709}
  (\bibinfo {year} {2011})}\BibitemShut {NoStop}%
\bibitem [{\citenamefont {Wehner}\ \emph {et~al.}(2018)\citenamefont {Wehner},
  \citenamefont {Elkouss},\ and\ \citenamefont {Hanson}}]{wehner_quantum_2018}%
  \BibitemOpen
  \bibfield  {author} {\bibinfo {author} {\bibfnamefont {S.}~\bibnamefont
  {Wehner}}, \bibinfo {author} {\bibfnamefont {D.}~\bibnamefont {Elkouss}}, \
  and\ \bibinfo {author} {\bibfnamefont {R.}~\bibnamefont {Hanson}},\
  }\href@noop {} {\bibfield  {journal} {\bibinfo  {journal} {Science}\ }\textbf
  {\bibinfo {volume} {362}},\ \bibinfo {pages} {eaam9288} (\bibinfo {year}
  {2018})}\BibitemShut {NoStop}%
\bibitem [{\citenamefont {Spring}\ \emph {et~al.}(2013)\citenamefont {Spring},
  \citenamefont {Metcalf}, \citenamefont {Humphreys}, \citenamefont
  {Kolthammer}, \citenamefont {Jin}, \citenamefont {Barbieri}, \citenamefont
  {Datta}, \citenamefont {Thomas-Peter}, \citenamefont {Langford},
  \citenamefont {Kundys} \emph {et~al.}}]{spring_quantum_2013}%
  \BibitemOpen
  \bibfield  {author} {\bibinfo {author} {\bibfnamefont {J.~B.}\ \bibnamefont
  {Spring}}, \bibinfo {author} {\bibfnamefont {B.~J.}\ \bibnamefont {Metcalf}},
  \bibinfo {author} {\bibfnamefont {P.~C.}\ \bibnamefont {Humphreys}}, \bibinfo
  {author} {\bibfnamefont {W.~S.}\ \bibnamefont {Kolthammer}}, \bibinfo
  {author} {\bibfnamefont {X.-M.}\ \bibnamefont {Jin}}, \bibinfo {author}
  {\bibfnamefont {M.}~\bibnamefont {Barbieri}}, \bibinfo {author}
  {\bibfnamefont {A.}~\bibnamefont {Datta}}, \bibinfo {author} {\bibfnamefont
  {N.}~\bibnamefont {Thomas-Peter}}, \bibinfo {author} {\bibfnamefont {N.~K.}\
  \bibnamefont {Langford}}, \bibinfo {author} {\bibfnamefont {D.}~\bibnamefont
  {Kundys}},  \emph {et~al.},\ }\href@noop {} {\bibfield  {journal} {\bibinfo
  {journal} {Science}\ }\textbf {\bibinfo {volume} {339}},\ \bibinfo {pages}
  {798} (\bibinfo {year} {2013})}\BibitemShut {NoStop}%
\bibitem [{\citenamefont {Li}\ \emph {et~al.}(2016)\citenamefont {Li},
  \citenamefont {Zhou}, \citenamefont {Rasmita}, \citenamefont {Aharonovich},\
  and\ \citenamefont {Gao}}]{li_nonblinking_2016}%
  \BibitemOpen
  \bibfield  {author} {\bibinfo {author} {\bibfnamefont {K.}~\bibnamefont
  {Li}}, \bibinfo {author} {\bibfnamefont {Y.}~\bibnamefont {Zhou}}, \bibinfo
  {author} {\bibfnamefont {A.}~\bibnamefont {Rasmita}}, \bibinfo {author}
  {\bibfnamefont {I.}~\bibnamefont {Aharonovich}}, \ and\ \bibinfo {author}
  {\bibfnamefont {W.}~\bibnamefont {Gao}},\ }\href@noop {} {\bibfield
  {journal} {\bibinfo  {journal} {Phys. Rev. Appl.}\ }\textbf {\bibinfo
  {volume} {6}},\ \bibinfo {pages} {024010} (\bibinfo {year}
  {2016})}\BibitemShut {NoStop}%
\bibitem [{\citenamefont {Zhou}\ \emph {et~al.}(2017)\citenamefont {Zhou},
  \citenamefont {Rasmita}, \citenamefont {Li}, \citenamefont {Xiong},
  \citenamefont {Aharonovich},\ and\ \citenamefont {Gao}}]{zhou_coherent_2017}%
  \BibitemOpen
  \bibfield  {author} {\bibinfo {author} {\bibfnamefont {Y.}~\bibnamefont
  {Zhou}}, \bibinfo {author} {\bibfnamefont {A.}~\bibnamefont {Rasmita}},
  \bibinfo {author} {\bibfnamefont {K.}~\bibnamefont {Li}}, \bibinfo {author}
  {\bibfnamefont {Q.}~\bibnamefont {Xiong}}, \bibinfo {author} {\bibfnamefont
  {I.}~\bibnamefont {Aharonovich}}, \ and\ \bibinfo {author} {\bibfnamefont
  {W.-b.}\ \bibnamefont {Gao}},\ }\href@noop {} {\bibfield  {journal} {\bibinfo
   {journal} {Nat. Commun.}\ }\textbf {\bibinfo {volume} {8}},\ \bibinfo
  {pages} {1} (\bibinfo {year} {2017})}\BibitemShut {NoStop}%
\bibitem [{\citenamefont {Oskooi}\ \emph {et~al.}(2010)\citenamefont {Oskooi},
  \citenamefont {Roundy}, \citenamefont {Ibanescu}, \citenamefont {Bermel},
  \citenamefont {Joannopoulos},\ and\ \citenamefont
  {Johnson}}]{oskooi_meep_2010}%
  \BibitemOpen
  \bibfield  {author} {\bibinfo {author} {\bibfnamefont {A.~F.}\ \bibnamefont
  {Oskooi}}, \bibinfo {author} {\bibfnamefont {D.}~\bibnamefont {Roundy}},
  \bibinfo {author} {\bibfnamefont {M.}~\bibnamefont {Ibanescu}}, \bibinfo
  {author} {\bibfnamefont {P.}~\bibnamefont {Bermel}}, \bibinfo {author}
  {\bibfnamefont {J.~D.}\ \bibnamefont {Joannopoulos}}, \ and\ \bibinfo
  {author} {\bibfnamefont {S.~G.}\ \bibnamefont {Johnson}},\ }\href@noop {}
  {\bibfield  {journal} {\bibinfo  {journal} {Comput. Phys. Commun.}\ }\textbf
  {\bibinfo {volume} {181}},\ \bibinfo {pages} {687} (\bibinfo {year}
  {2010})}\BibitemShut {NoStop}%
\bibitem [{\citenamefont {Akahane}\ \emph {et~al.}(2003)\citenamefont
  {Akahane}, \citenamefont {Asano}, \citenamefont {Song},\ and\ \citenamefont
  {Noda}}]{akahane_high_2003}%
  \BibitemOpen
  \bibfield  {author} {\bibinfo {author} {\bibfnamefont {Y.}~\bibnamefont
  {Akahane}}, \bibinfo {author} {\bibfnamefont {T.}~\bibnamefont {Asano}},
  \bibinfo {author} {\bibfnamefont {B.-S.}\ \bibnamefont {Song}}, \ and\
  \bibinfo {author} {\bibfnamefont {S.}~\bibnamefont {Noda}},\ }\href@noop {}
  {\bibfield  {journal} {\bibinfo  {journal} {Nature}\ }\textbf {\bibinfo
  {volume} {425}},\ \bibinfo {pages} {944} (\bibinfo {year}
  {2003})}\BibitemShut {NoStop}%
\bibitem [{\citenamefont {Purcell}\ \emph {et~al.}(1946)\citenamefont
  {Purcell}, \citenamefont {Torrey},\ and\ \citenamefont
  {Pound}}]{purcell_resonance_1946}%
  \BibitemOpen
  \bibfield  {author} {\bibinfo {author} {\bibfnamefont {E.~M.}\ \bibnamefont
  {Purcell}}, \bibinfo {author} {\bibfnamefont {H.~C.}\ \bibnamefont {Torrey}},
  \ and\ \bibinfo {author} {\bibfnamefont {R.~V.}\ \bibnamefont {Pound}},\
  }\href@noop {} {\bibfield  {journal} {\bibinfo  {journal} {Phys. Rev.}\
  }\textbf {\bibinfo {volume} {69}},\ \bibinfo {pages} {37} (\bibinfo {year}
  {1946})}\BibitemShut {NoStop}%
\bibitem [{\citenamefont {Taflove}\ \emph {et~al.}(2013)\citenamefont
  {Taflove}, \citenamefont {Oskooi},\ and\ \citenamefont
  {Johnson}}]{taflove_photonics_2013}%
  \BibitemOpen
  \bibfield  {author} {\bibinfo {author} {\bibfnamefont {A.}~\bibnamefont
  {Taflove}}, \bibinfo {author} {\bibfnamefont {A.}~\bibnamefont {Oskooi}}, \
  and\ \bibinfo {author} {\bibfnamefont {S.~G.}\ \bibnamefont {Johnson}},\
  }\href@noop {} {\emph {\bibinfo {title} {Advances in FDTD computational
  electrodynamics: photonics and nanotechnology}}}\ (\bibinfo  {publisher}
  {Artech house},\ \bibinfo {year} {2013})\BibitemShut {NoStop}%
\bibitem [{\citenamefont {Zhang}\ \emph {et~al.}(2018)\citenamefont {Zhang},
  \citenamefont {Sun}, \citenamefont {Burek}, \citenamefont {Dory},
  \citenamefont {Tzeng}, \citenamefont {Fischer}, \citenamefont {Kelaita},
  \citenamefont {Lagoudakis}, \citenamefont {Radulaski}, \citenamefont {Shen},
  \citenamefont {Melosh}, \citenamefont {Chu}, \citenamefont {Loncar},\ and\
  \citenamefont {Vuckovic}}]{zhang_strongly_2018}%
  \BibitemOpen
  \bibfield  {author} {\bibinfo {author} {\bibfnamefont {J.~L.}\ \bibnamefont
  {Zhang}}, \bibinfo {author} {\bibfnamefont {S.}~\bibnamefont {Sun}}, \bibinfo
  {author} {\bibfnamefont {M.~J.}\ \bibnamefont {Burek}}, \bibinfo {author}
  {\bibfnamefont {C.}~\bibnamefont {Dory}}, \bibinfo {author} {\bibfnamefont
  {Y.-K.}\ \bibnamefont {Tzeng}}, \bibinfo {author} {\bibfnamefont {K.~A.}\
  \bibnamefont {Fischer}}, \bibinfo {author} {\bibfnamefont {Y.}~\bibnamefont
  {Kelaita}}, \bibinfo {author} {\bibfnamefont {K.~G.}\ \bibnamefont
  {Lagoudakis}}, \bibinfo {author} {\bibfnamefont {M.}~\bibnamefont
  {Radulaski}}, \bibinfo {author} {\bibfnamefont {Z.-X.}\ \bibnamefont {Shen}},
  \bibinfo {author} {\bibfnamefont {N.~A.}\ \bibnamefont {Melosh}}, \bibinfo
  {author} {\bibfnamefont {S.}~\bibnamefont {Chu}}, \bibinfo {author}
  {\bibfnamefont {M.}~\bibnamefont {Loncar}}, \ and\ \bibinfo {author}
  {\bibfnamefont {J.}~\bibnamefont {Vuckovic}},\ }\href {\doibase
  10.1021/acs.nanolett.7b05075} {\bibfield  {journal} {\bibinfo  {journal}
  {Nano Lett.}\ }\textbf {\bibinfo {volume} {18}},\ \bibinfo {pages} {1360}
  (\bibinfo {year} {2018})}\BibitemShut {NoStop}%
\bibitem [{\citenamefont {Meesala}\ \emph {et~al.}(2018)\citenamefont
  {Meesala}, \citenamefont {Sohn}, \citenamefont {Pingault}, \citenamefont
  {Shao}, \citenamefont {Atikian}, \citenamefont {Holzgrafe}, \citenamefont
  {G{\"u}ndo{\u{g}}an}, \citenamefont {Stavrakas}, \citenamefont {Sipahigil},
  \citenamefont {Chia} \emph {et~al.}}]{meesala_strain_2018}%
  \BibitemOpen
  \bibfield  {author} {\bibinfo {author} {\bibfnamefont {S.}~\bibnamefont
  {Meesala}}, \bibinfo {author} {\bibfnamefont {Y.-I.}\ \bibnamefont {Sohn}},
  \bibinfo {author} {\bibfnamefont {B.}~\bibnamefont {Pingault}}, \bibinfo
  {author} {\bibfnamefont {L.}~\bibnamefont {Shao}}, \bibinfo {author}
  {\bibfnamefont {H.~A.}\ \bibnamefont {Atikian}}, \bibinfo {author}
  {\bibfnamefont {J.}~\bibnamefont {Holzgrafe}}, \bibinfo {author}
  {\bibfnamefont {M.}~\bibnamefont {G{\"u}ndo{\u{g}}an}}, \bibinfo {author}
  {\bibfnamefont {C.}~\bibnamefont {Stavrakas}}, \bibinfo {author}
  {\bibfnamefont {A.}~\bibnamefont {Sipahigil}}, \bibinfo {author}
  {\bibfnamefont {C.}~\bibnamefont {Chia}},  \emph {et~al.},\ }\href@noop {}
  {\bibfield  {journal} {\bibinfo  {journal} {Phys. Rev. B}\ }\textbf {\bibinfo
  {volume} {97}},\ \bibinfo {pages} {205444} (\bibinfo {year}
  {2018})}\BibitemShut {NoStop}%
\bibitem [{\citenamefont {Rogers}\ \emph {et~al.}(2014)\citenamefont {Rogers},
  \citenamefont {Jahnke}, \citenamefont {Metsch}, \citenamefont {Sipahigil},
  \citenamefont {Binder}, \citenamefont {Teraji}, \citenamefont {Sumiya},
  \citenamefont {Isoya}, \citenamefont {Lukin}, \citenamefont {Hemmer},\ and\
  \citenamefont {Jelezko}}]{rogers_all-optical_2014}%
  \BibitemOpen
  \bibfield  {author} {\bibinfo {author} {\bibfnamefont {L.~J.}\ \bibnamefont
  {Rogers}}, \bibinfo {author} {\bibfnamefont {K.~D.}\ \bibnamefont {Jahnke}},
  \bibinfo {author} {\bibfnamefont {M.~H.}\ \bibnamefont {Metsch}}, \bibinfo
  {author} {\bibfnamefont {A.}~\bibnamefont {Sipahigil}}, \bibinfo {author}
  {\bibfnamefont {J.~M.}\ \bibnamefont {Binder}}, \bibinfo {author}
  {\bibfnamefont {T.}~\bibnamefont {Teraji}}, \bibinfo {author} {\bibfnamefont
  {H.}~\bibnamefont {Sumiya}}, \bibinfo {author} {\bibfnamefont
  {J.}~\bibnamefont {Isoya}}, \bibinfo {author} {\bibfnamefont {M.~D.}\
  \bibnamefont {Lukin}}, \bibinfo {author} {\bibfnamefont {P.}~\bibnamefont
  {Hemmer}}, \ and\ \bibinfo {author} {\bibfnamefont {F.}~\bibnamefont
  {Jelezko}},\ }\href {\doibase 10.1103/PhysRevLett.113.263602} {\bibfield
  {journal} {\bibinfo  {journal} {Phys. Rev. Lett.}\ }\textbf {\bibinfo
  {volume} {113}},\ \bibinfo {pages} {263602} (\bibinfo {year}
  {2014})}\BibitemShut {NoStop}%
\bibitem [{\citenamefont {Hepp}\ \emph {et~al.}(2014)\citenamefont {Hepp},
  \citenamefont {M{\"u}ller}, \citenamefont {Waselowski}, \citenamefont
  {Becker}, \citenamefont {Pingault}, \citenamefont {Sternschulte},
  \citenamefont {Steinm{\"u}ller-Nethl}, \citenamefont {Gali}, \citenamefont
  {Maze}, \citenamefont {Atat{\"u}re} \emph {et~al.}}]{hepp_electronic_2014}%
  \BibitemOpen
  \bibfield  {author} {\bibinfo {author} {\bibfnamefont {C.}~\bibnamefont
  {Hepp}}, \bibinfo {author} {\bibfnamefont {T.}~\bibnamefont {M{\"u}ller}},
  \bibinfo {author} {\bibfnamefont {V.}~\bibnamefont {Waselowski}}, \bibinfo
  {author} {\bibfnamefont {J.~N.}\ \bibnamefont {Becker}}, \bibinfo {author}
  {\bibfnamefont {B.}~\bibnamefont {Pingault}}, \bibinfo {author}
  {\bibfnamefont {H.}~\bibnamefont {Sternschulte}}, \bibinfo {author}
  {\bibfnamefont {D.}~\bibnamefont {Steinm{\"u}ller-Nethl}}, \bibinfo {author}
  {\bibfnamefont {A.}~\bibnamefont {Gali}}, \bibinfo {author} {\bibfnamefont
  {J.~R.}\ \bibnamefont {Maze}}, \bibinfo {author} {\bibfnamefont
  {M.}~\bibnamefont {Atat{\"u}re}},  \emph {et~al.},\ }\href@noop {} {\bibfield
   {journal} {\bibinfo  {journal} {Phys. Rev. Lett.}\ }\textbf {\bibinfo
  {volume} {112}},\ \bibinfo {pages} {036405} (\bibinfo {year}
  {2014})}\BibitemShut {NoStop}%
\bibitem [{\citenamefont {Imamoglu}\ \emph {et~al.}(1999)\citenamefont
  {Imamoglu}, \citenamefont {Awschalom}, \citenamefont {Burkard}, \citenamefont
  {DiVincenzo}, \citenamefont {Loss}, \citenamefont {Sherwin},\ and\
  \citenamefont {Small}}]{imamoglu_quantum_1999}%
  \BibitemOpen
  \bibfield  {author} {\bibinfo {author} {\bibfnamefont {A.}~\bibnamefont
  {Imamoglu}}, \bibinfo {author} {\bibfnamefont {D.~D.}\ \bibnamefont
  {Awschalom}}, \bibinfo {author} {\bibfnamefont {G.}~\bibnamefont {Burkard}},
  \bibinfo {author} {\bibfnamefont {D.~P.}\ \bibnamefont {DiVincenzo}},
  \bibinfo {author} {\bibfnamefont {D.}~\bibnamefont {Loss}}, \bibinfo {author}
  {\bibfnamefont {M.}~\bibnamefont {Sherwin}}, \ and\ \bibinfo {author}
  {\bibfnamefont {A.}~\bibnamefont {Small}},\ }\href {\doibase
  10.1103/PhysRevLett.83.4204} {\bibfield  {journal} {\bibinfo  {journal}
  {Phys. Rev. Lett.}\ }\textbf {\bibinfo {volume} {83}},\ \bibinfo {pages}
  {4204} (\bibinfo {year} {1999})}\BibitemShut {NoStop}%
\bibitem [{\citenamefont {Kastoryano}\ \emph {et~al.}(2011)\citenamefont
  {Kastoryano}, \citenamefont {Reiter},\ and\ \citenamefont
  {Sørensen}}]{kastoryano_dissipative_2011}%
  \BibitemOpen
  \bibfield  {author} {\bibinfo {author} {\bibfnamefont {M.~J.}\ \bibnamefont
  {Kastoryano}}, \bibinfo {author} {\bibfnamefont {F.}~\bibnamefont {Reiter}},
  \ and\ \bibinfo {author} {\bibfnamefont {A.~S.}\ \bibnamefont {Sørensen}},\
  }\href {\doibase 10.1103/PhysRevLett.106.090502} {\bibfield  {journal}
  {\bibinfo  {journal} {Phys. Rev. Lett.}\ }\textbf {\bibinfo {volume} {106}},\
  \bibinfo {pages} {090502} (\bibinfo {year} {2011})}\BibitemShut {NoStop}%
\bibitem [{\citenamefont {Zheng}\ and\ \citenamefont
  {Guo}(2000)}]{zheng_efficient_2000}%
  \BibitemOpen
  \bibfield  {author} {\bibinfo {author} {\bibfnamefont {S.-B.}\ \bibnamefont
  {Zheng}}\ and\ \bibinfo {author} {\bibfnamefont {G.-C.}\ \bibnamefont
  {Guo}},\ }\href {\doibase 10.1103/PhysRevLett.85.2392} {\bibfield  {journal}
  {\bibinfo  {journal} {Phys. Rev. Lett.}\ }\textbf {\bibinfo {volume} {85}},\
  \bibinfo {pages} {2392} (\bibinfo {year} {2000})}\BibitemShut {NoStop}%
\bibitem [{\citenamefont {Waks}\ and\ \citenamefont
  {Vuckovic}(2006)}]{waks_dipole_2006}%
  \BibitemOpen
  \bibfield  {author} {\bibinfo {author} {\bibfnamefont {E.}~\bibnamefont
  {Waks}}\ and\ \bibinfo {author} {\bibfnamefont {J.}~\bibnamefont
  {Vuckovic}},\ }\href {\doibase 10.1103/PhysRevLett.96.153601} {\bibfield
  {journal} {\bibinfo  {journal} {Phys. Rev. Lett.}\ }\textbf {\bibinfo
  {volume} {96}},\ \bibinfo {pages} {153601} (\bibinfo {year}
  {2006})}\BibitemShut {NoStop}%
\bibitem [{\citenamefont {Borregaard}\ \emph {et~al.}(2015)\citenamefont
  {Borregaard}, \citenamefont {Kómár}, \citenamefont {Kessler}, \citenamefont
  {Lukin},\ and\ \citenamefont {Sørensen}}]{borregaard_long-distance_2015}%
  \BibitemOpen
  \bibfield  {author} {\bibinfo {author} {\bibfnamefont {J.}~\bibnamefont
  {Borregaard}}, \bibinfo {author} {\bibfnamefont {P.}~\bibnamefont {Kómár}},
  \bibinfo {author} {\bibfnamefont {E.~M.}\ \bibnamefont {Kessler}}, \bibinfo
  {author} {\bibfnamefont {M.~D.}\ \bibnamefont {Lukin}}, \ and\ \bibinfo
  {author} {\bibfnamefont {A.~S.}\ \bibnamefont {Sørensen}},\ }\href {\doibase
  10.1103/PhysRevA.92.012307} {\bibfield  {journal} {\bibinfo  {journal} {Phys.
  Rev. A}\ }\textbf {\bibinfo {volume} {92}},\ \bibinfo {pages} {012307}
  (\bibinfo {year} {2015})}\BibitemShut {NoStop}%
\bibitem [{\citenamefont {Cirac}\ \emph {et~al.}(1997)\citenamefont {Cirac},
  \citenamefont {Zoller}, \citenamefont {Kimble},\ and\ \citenamefont
  {Mabuchi}}]{cirac_quantum_1997}%
  \BibitemOpen
  \bibfield  {author} {\bibinfo {author} {\bibfnamefont {J.~I.}\ \bibnamefont
  {Cirac}}, \bibinfo {author} {\bibfnamefont {P.}~\bibnamefont {Zoller}},
  \bibinfo {author} {\bibfnamefont {H.~J.}\ \bibnamefont {Kimble}}, \ and\
  \bibinfo {author} {\bibfnamefont {H.}~\bibnamefont {Mabuchi}},\ }\href
  {\doibase 10.1103/PhysRevLett.78.3221} {\bibfield  {journal} {\bibinfo
  {journal} {Phys. Rev. Lett.}\ }\textbf {\bibinfo {volume} {78}},\ \bibinfo
  {pages} {3221} (\bibinfo {year} {1997})}\BibitemShut {NoStop}%
\bibitem [{\citenamefont {Kok}\ \emph {et~al.}(2007)\citenamefont {Kok},
  \citenamefont {Munro}, \citenamefont {Nemoto}, \citenamefont {Ralph},
  \citenamefont {Dowling},\ and\ \citenamefont
  {Milburn}}]{kok_publishers_2007}%
  \BibitemOpen
  \bibfield  {author} {\bibinfo {author} {\bibfnamefont {P.}~\bibnamefont
  {Kok}}, \bibinfo {author} {\bibfnamefont {W.~J.}\ \bibnamefont {Munro}},
  \bibinfo {author} {\bibfnamefont {K.}~\bibnamefont {Nemoto}}, \bibinfo
  {author} {\bibfnamefont {T.~C.}\ \bibnamefont {Ralph}}, \bibinfo {author}
  {\bibfnamefont {J.~P.}\ \bibnamefont {Dowling}}, \ and\ \bibinfo {author}
  {\bibfnamefont {G.~J.}\ \bibnamefont {Milburn}},\ }\href@noop {} {\bibfield
  {journal} {\bibinfo  {journal} {Rev. Mod. Phys.}\ }\textbf {\bibinfo {volume}
  {79}},\ \bibinfo {pages} {135} (\bibinfo {year} {2007})}\BibitemShut
  {NoStop}%
\bibitem [{\citenamefont {Riedmatten}\ and\ \citenamefont
  {Afzelius}(2015)}]{de_riedmatten_quantum_2015}%
  \BibitemOpen
  \bibfield  {author} {\bibinfo {author} {\bibfnamefont {H.}~\bibnamefont
  {Riedmatten}}\ and\ \bibinfo {author} {\bibfnamefont {M.}~\bibnamefont
  {Afzelius}},\ }\href {\doibase 10.1007/978-3-319-19231-4_9} {\emph {\bibinfo
  {title} {Engineering the Atom-Photon Interaction: Controlling Fundamental
  Processes with Photons, Atoms and Solids}}},\ edited by\ \bibinfo {editor}
  {\bibfnamefont {A.}~\bibnamefont {Predojević}}\ and\ \bibinfo {editor}
  {\bibfnamefont {M.~W.}\ \bibnamefont {Mitchell}},\ Nano-Optics and
  Nanophotonics\ (\bibinfo  {publisher} {Springer International Publishing},\
  \bibinfo {year} {2015})\ pp.\ \bibinfo {pages} {241--273}\BibitemShut
  {NoStop}%
\bibitem [{\citenamefont {Munro}\ \emph {et~al.}(2015)\citenamefont {Munro},
  \citenamefont {Azuma}, \citenamefont {Tamaki},\ and\ \citenamefont
  {Nemoto}}]{munro_inside_2015}%
  \BibitemOpen
  \bibfield  {author} {\bibinfo {author} {\bibfnamefont {W.~J.}\ \bibnamefont
  {Munro}}, \bibinfo {author} {\bibfnamefont {K.}~\bibnamefont {Azuma}},
  \bibinfo {author} {\bibfnamefont {K.}~\bibnamefont {Tamaki}}, \ and\ \bibinfo
  {author} {\bibfnamefont {K.}~\bibnamefont {Nemoto}},\ }\href {\doibase
  10.1109/JSTQE.2015.2392076} {\bibfield  {journal} {\bibinfo  {journal} {IEEE
  J. Sel. Top. Quantum Electron.}\ }\textbf {\bibinfo {volume} {21}},\ \bibinfo
  {pages} {78} (\bibinfo {year} {2015})}\BibitemShut {NoStop}%
\bibitem [{\citenamefont {Terhal}(2015)}]{terhal_quantum_2015}%
  \BibitemOpen
  \bibfield  {author} {\bibinfo {author} {\bibfnamefont {B.~M.}\ \bibnamefont
  {Terhal}},\ }\href {\doibase 10.1103/RevModPhys.87.307} {\bibfield  {journal}
  {\bibinfo  {journal} {Rev. Mod. Phys.}\ }\textbf {\bibinfo {volume} {87}},\
  \bibinfo {pages} {307} (\bibinfo {year} {2015})}\BibitemShut {NoStop}%
\bibitem [{\citenamefont {Degen}\ \emph {et~al.}(2017)\citenamefont {Degen},
  \citenamefont {Reinhard},\ and\ \citenamefont
  {Cappellaro}}]{degen_quantum_2017}%
  \BibitemOpen
  \bibfield  {author} {\bibinfo {author} {\bibfnamefont {C.~L.}\ \bibnamefont
  {Degen}}, \bibinfo {author} {\bibfnamefont {F.}~\bibnamefont {Reinhard}}, \
  and\ \bibinfo {author} {\bibfnamefont {P.}~\bibnamefont {Cappellaro}},\
  }\href {\doibase 10.1103/RevModPhys.89.035002} {\bibfield  {journal}
  {\bibinfo  {journal} {Rev. Mod. Phys.}\ }\textbf {\bibinfo {volume} {89}},\
  \bibinfo {pages} {035002} (\bibinfo {year} {2017})}\BibitemShut {NoStop}%
\end{thebibliography}
\end{document}